\documentclass[%
reprint,
superscriptaddress,
nofootinbib,
longbibliography,
amsmath,
amssymb,
aps,
showkeys
]{revtex4-2}
\usepackage{graphicx}
\usepackage{dcolumn}
\usepackage{bm}
\usepackage{hyperref}
\usepackage[bottom]{footmisc}
\usepackage{footnote}
\usepackage{xcolor}
\usepackage{booktabs}
\usepackage{array,multirow}
\usepackage{amsmath}
\usepackage{amsfonts}
\usepackage{amssymb}
\usepackage{mathrsfs}
\usepackage{enumerate}
\usepackage{float}
\usepackage{verbatim}
\usepackage{braket}
\usepackage{setspace}
\usepackage[normalem]{ulem}
\usepackage{multirow}
\usepackage{comment}
\allowdisplaybreaks
\hypersetup{breaklinks=true, colorlinks=true, citecolor=blue, linkcolor=blue, urlcolor=blue}

\begin{document}

\newcommand{\LAL}{$\Lambda\bar{\Lambda}$}
\newcommand{\PLal}{P_{\Lambda\bar{\Lambda}}}
\newcommand{\DR}{\Delta R}
\newcommand{\Dy}{\Delta y}
\newcommand{\Dphi}{\Delta\phi}
\newcommand{\sqs}{\sqrt{s}}
\newcommand{\dNdy}{\mathrm{d}N/\mathrm{d}y}
\newcommand{\Npair}{N_\mathrm{pair}}
\newcommand{\chisq}{\chi^2}
\newcommand{\ffd}{f_\mathrm{d}}


\title{Quantum decoherence of hyperon spin correlations in QCD hadronization }

\author{Feng Liu}
\email{feng.liu.2@stonybrook.edu}
\affiliation{Department of Physics and Astronomy, Stony Brook University,
             New York 11794-3800, USA}

\author{Zhoudunming~Tu}
\email{zhoudunming@bnl.gov}
\affiliation{Department of Physics and Astronomy, Stony Brook University,
             New York 11794-3800, USA}
\affiliation{Department of Physics, Brookhaven National Laboratory,
             Upton, New York 11973-5000, USA}

\date{\today}

\begin{abstract}
Hadronization, the transition of quarks and gluons into hadrons, lies beyond the reach of perturbative quantum chromodynamics (QCD) and is commonly described by the semiclassical Lund string model. Yet this very success raises a fundamental question: where does the \textit{quantumness} go during hadronization? In this Letter, we propose an approach inspired by quantum information science, in which (i) quark-antiquark pairs excited from the QCD vacuum inherit its quantum numbers, giving rise to spin entanglement at their creation, and (ii) subsequent string breaking generates environmental degrees of freedom that induce quantum decoherence of the spin state. This framework simultaneously describes the $\Lambda$ hyperon spin-correlation data measured at RHIC [\hyperlink{https://www.nature.com/articles/s41586-025-09920-0}{Nature 650, 65--71 (2026)}] and at the LHC, establishing a quantitative connection between the QCD vacuum, spin entanglement and decoherence, and hadronization.
\end{abstract}

\maketitle
\textit{\textbf{Introduction.}}~Color confinement in quantum chromodynamics (QCD) dictates that quarks and gluons produced in high-energy collisions must materialize as colorless hadrons. This conversion, known as hadronization, is intrinsically nonperturbative, and its description still relies largely on phenomenological models~\cite{Andersson:1983ia,Webber:1983if}. 

Among the most successful is the Lund string model~\cite{Andersson:1983ia}, in which hadrons form through the breaking of color flux tubes (strings) stretched between colored partons: the string breaks successively via a Schwinger-like mechanism of pair production from the vacuum until its stored energy is exhausted and the produced partons bind into hadrons. The Lund model is essentially semiclassical, it treats hadronization as a stochastic process without retaining quantum phase information, yet it successfully describes a broad class of observables~\cite{Sjostrand:1987xj,Andersson:1997xwk,Skands:2014pea}. This raises a natural question: \textit{how does such a classical probabilistic description emerge from the underlying quantum dynamics of hadronization?}

In recent years, concepts and tools from quantum information science have opened new avenues for exploring the quantum nature of particle systems~\cite{Kharzeev:2021nzh,Florio:2023dke,Hentschinski:2024gaa,ATLAS:2023fsd,CMS:2024pts,PhysRevLett.134.111902,Hentschinski:2023izh,Tu:2019ouv,Kharzeev:2026jkq,Grieninger:2025rdi,ATLAS:2026hye,Gu:2025ijz,Aoude:2025ovu,Gong:2021bcp,Barata:2023jgd,Amorosso:2024leg,Amorosso:2024glf,Amorosso:2026mdo,Hatta:2024lbw,Bhattacharya:2024sno,Fucilla:2025kit,Hatta:2025obw}, offering a fresh perspective on this question. In particular, simulations of string-breaking dynamics in the (1+1)-dimensional Schwinger model~\cite{Grieninger:2025rdi} have revealed that strong quantum entanglement emerges in a nearly breaking color string, suggesting that the color flux tube should be viewed as a highly entangled quantum state rather than a purely classical string. In such a system, a subsystem loses quantum phase information once the unobserved degrees of freedom are traced out, leading to quantum decoherence~\cite{Schlosshauer:2019ewh,Aoude:2025ovu,Gu:2025ijz,Lin:2025eci}. A notable consequence is that a subsystem of the string exhibits effectively classical behavior even though the full string remains highly entangled---providing a new understanding of the question raised above.

\label{sec:decoherence}
 \begin{figure*}[t]
    \centering
    \includegraphics[width=0.99\textwidth]{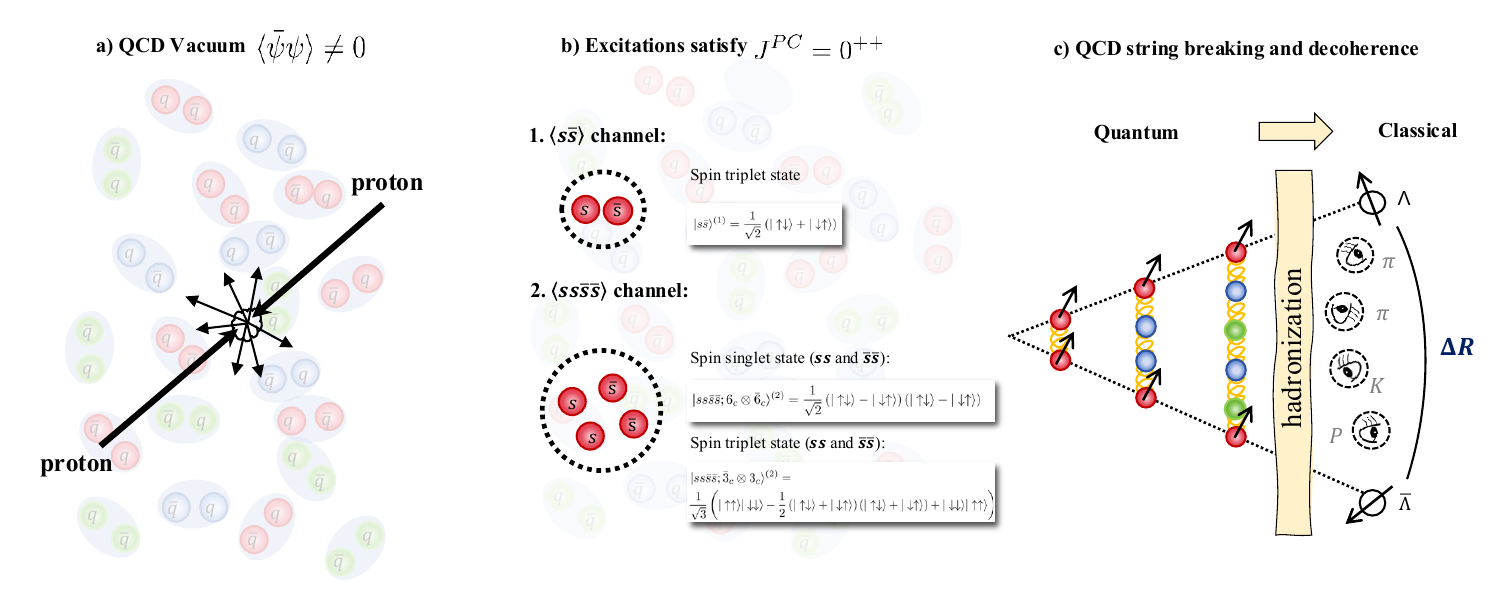}
    \caption{Illustration of spin entanglement of quark-antiquark pairs in the QCD vacuum and its hadronization process.}
    \label{fig:fig1}
\end{figure*}

On the experimental side, few measurements had been identified as sensitive to this quantum perspective of hadronization until the recent discovery of $\Lambda$ hyperon spin correlations by the STAR collaboration~\cite{STAR:2025njp}. Notably, such spin correlations can be traced back to the strange quark-antiquark pairs ($s\bar{s}$) produced from the QCD vacuum~\cite{Ellis:1995fc}. Moreover, the observed spin correlation exhibits a clear dependence on the relative angular separation $\Delta R$ between the $\Lambda$ hyperons, suggesting the presence of quantum decoherence during hadronization. The $\Delta R$ dependence thus provides a direct experimental handle on the quantum-to-classical transition during hadronization.

In this Letter, we introduce a phenomenological framework that describes how the spin state of quark-antiquark pairs is determined at their creation from the QCD vacuum and how the resulting quantum correlations evolve during hadronization. Specifically, we propose that the observed $\Lambda$ hyperon spin correlations arise as follows:
\begin{enumerate}[a)]
    \item the QCD vacuum exhibits a non-vanishing quark condensate due to the spontaneous chiral symmetry breaking, $\langle\bar{\psi}\psi\rangle \neq 0$; high-energy collisions can excite these virtual quark-antiquark pairs and convert them into real particles.  
    \item The $s\bar{s}$ pairs created from the QCD vacuum inherit the vacuum quantum numbers, $J^{PC}=0^{++}$, which gives rise to the spin correlation within each pair; for a single $s\bar{s}$ pair, they are in a maximally entangled state. We denote this part as the \textit{Vacuum Spin Chain(VSC)} model.
    \item Hadronization subsequently proceeds via string breaking, with additional quarks excited in between; these intermediate quarks act as environmental degrees of freedom that monitor the $s\bar{s}$ pairs, leading to decoherence of their spin correlations and the emergence of a quantum-to-classical transition. We denote the decoherence as ``\textit{the Witness effect}".
\end{enumerate}
The process is illustrated in Fig.~\ref{fig:fig1}.

Within this framework, the model simultaneously describes the published STAR results at 200~GeV and the preliminary $\Lambda\bar{\Lambda}$ and $\Lambda\Lambda$ results at 13~TeV in proton-proton collisions. This constitutes a first-of-its-kind study relating quantum decoherence to hadronization quantitatively through measured data.

\textit{\textbf{The VSC model.}}~The simplest vacuum excitation channel is the production of a single $q\bar{q}$ pair. However, the vacuum may also be excited coherently into multiple pairs; in particular, the production of two $q\bar{q}$ pairs has been used to account for baryon formation in the Lund string model~\cite{Andersson:1981ce,Andersson:1984af,Casher:1978wy,TPCTwoGamma:1985zxy}. To simultaneously account for the spin correlations of $\Lambda\bar{\Lambda}$ and $\Lambda\Lambda$~($\bar{\Lambda}\bar{\Lambda}$) pairs, the present work considers $s\bar{s}$ and $ss\bar{s}\bar{s}$ as one-pair and two-pair channels, respectively. 

\textit{One-pair channel.}~We begin with one pair of $s\bar{s}$ carrying $J^{PC}=0^{++}$ excited from vacuum. Here the $J$, $P$ and $C$ represent the total angular momentum, parity, and charge conjugation, respectively. The quantum state is denoted as $\ket{s\bar{s}}^{(1)}$, which can be expressed as 
\begin{equation}
\left.
\begin{array}{l}
J = L + S, L + S - 1, \dots, |L - S| \\
P = \eta_s \eta_{\bar{s}} (-1)^{L} \\
C = (-1)^{L+S}
\end{array}
\right\}
\end{equation}
where the $\eta_{s(\bar{s})}$ is the intrinsic parity of $s$~$(\bar{s})$ quark. Requiring the pair $\ket{s\bar{s}}^{(1)}$ to carry $J^{PC}=0^{++}$ uniquely leads to one solution, i.e., $L=S=1$~\cite{LeYaouanc:1972vsx}. 

Based on this solution, we can write the $L$ and $S$ wavefunction explicitly as,  
\begin{align}
&~~~~~~~~\ket{J=0} =  \frac{1}{\sqrt{3}} \ket{L_z=1}\ket{S_z=-1} \nonumber\\
  &  - \frac{1}{\sqrt{3}}\ket{L_z=0}\ket{S_z=0} + \frac{1}{\sqrt{3}}\ket{L_z=-1}\ket{S_z=1}.
  \label{eq:J0}
\end{align}
In the momentum representation, the orbital angular momentum states are realized by the spherical harmonics, i.e., $\braket{\theta,\phi|L_z=m} = Y_1^{m}(\theta,\phi)$, where $\theta$ and $\phi$ denote the polar and azimuthal angles of $s$ quark in the pair rest frame, respectively . The spin-triplet ($S=1$) states are constructed from the direct product of single-particle spin states as: $\ket{S_z=1} = \ket{\uparrow\uparrow}$, $\ket{S_z=0} = \frac{1}{\sqrt{2} } \left( \ket{\uparrow\downarrow} + \ket{\downarrow \uparrow} \right)$, and $\ket{S_z=-1} = \ket{\downarrow\downarrow}$. Substituting these expressions into Eq.~\ref{eq:J0} yields the wave function of the $s\bar{s}$ pair as, 
\begin{align}
      &~~~~~~~~\ket{s\bar{s}}^{(1)}=-\sqrt{\frac{1}{8\pi}} \sin\theta e^{i\phi}\ket{\downarrow\downarrow}
      \nonumber\\
   & -\sqrt{\frac{1}{8\pi}}  \cos \theta \left( \ket{\uparrow\downarrow}  + \ket{\downarrow\uparrow}  \right) + \sqrt{\frac{1}{8\pi}}  \sin \theta e^{-i\phi} \ket{\uparrow\uparrow}.
   \label{eq:J0wavefunction}
\end{align}
This wave function is identical to the Bell state~\cite{NielsenChuang} given by
\begin{equation}
    \ket{\psi^+,\hat{n}} = \frac{1}{\sqrt{2}}\left( \ket{\uparrow\downarrow,\hat{n}} + \ket{\downarrow \uparrow,\hat{n}}\right), 
\end{equation}
where the quantized axis is chosen as the direction of $s$ quark momentum $\hat{n}= (\sin \theta \cos \phi,\sin\theta \sin\phi,\cos\theta  )$. Therefore, rotating Eq.~\ref{eq:J0wavefunction} to the helicity frame of $s\bar{s}$ pair, defined by taking the $z$ axis along the $s$ quark momentum, the spin state can be written as
\begin{equation}
\ket{s\bar{s}}^{(1)} = \frac{1}{\sqrt{2}}\left( \ket{\uparrow\downarrow} + \ket{\downarrow\uparrow} \right),
\label{eq:ssbar}
\end{equation}
which exhibits \textit{maximal quantum entanglement}. 

The corresponding spin correlation between $s$ and $\bar{s}$  as defined in Ref.~\cite{STAR:2025njp} is 
\begin{equation}
    P_{s\bar{s}}^{(1)} = \frac{1}{3}.
\end{equation}

\textit{Two-pair channel.} We now extend the formulism to the case of two-pair channel, where a system of $ss\bar{s}\bar{s}$ carries vacuum quantum number $J^{PC}=0^{++}$. Note that the identical-particle symmetry requires the total wave function to be antisymmetric under exchange of the two identical $s$~($\bar{s}$) quarks. A convenient way to implement this constraint is to organize the full system into a diquark--antidiquark configuration, such as $ss$ and $\bar{s}\bar{s}$. 

The spin and color wave function must be correlated such that their product is antisymmetric under identical particle exchange. The detailed calculations are provided in the Appendix. Consequently, this gives rise to two allowed spin configurations.

Configuration (a): the state $\ket{ss\bar{s}\bar{s};6_c\otimes\bar{6}_c}$, where $ss~(\bar{s}\bar{s})$ is in the symmetric color-(anti) sextet state and in the spin singlet. The wave function is given by 
\begin{equation}
    \ket{ss\bar{s}\bar{s};6_c\otimes\bar{6}_c}^{(2)} = \frac{1}{2} \left( \ket{\uparrow\downarrow}-\ket{\downarrow\uparrow}\right)\left(\ket{\uparrow\downarrow} - \ket{\downarrow\uparrow} \right)
    \label{eq:color_symmetric}
\end{equation}
The spin correlations between  $ss$ ($\bar{s}\bar{s}$) and $s\bar{s}$ are $P^{(2)}_{ss(\bar{s}\bar{s}),(a)} = -1$ and $P^{(2)}_{s\bar{s},(a)} = 0$, respectively.

Configuration (b): the state $\ket{ss\bar{s}\bar{s};\bar{3}_c\otimes3_c}$, where $\bar{s}\bar{s}~(ss)$ is in the antisymmetric color-(anti) triplet state and in the spin triplet. The wave function is given by, 
\begin{multline}
\ket{ss\bar{s}\bar{s};\bar{3}_c\otimes3_c}^{(2)} = \frac{1}{\sqrt{3}} \ket{S=1}_{ss}\ket{S=-1}_{\bar{s}\bar{s}} \\
- \frac{1}{\sqrt{3}} \ket{S=0}_{ss}\ket{S=0}_{\bar{s}\bar{s}} + \frac{1}{\sqrt{3}} \ket{S=-1}_{ss}\ket{S=1}_{\bar{s}\bar{s}} \\
= \frac{1}{\sqrt{3}} \Big[ \ket{\uparrow\uparrow}\ket{\downarrow\downarrow} + \ket{\downarrow\downarrow}\ket{\uparrow\uparrow} \\
- \tfrac{1}{2} ( \ket{\uparrow\downarrow} + \ket{\downarrow\uparrow} ) ( \ket{\uparrow\downarrow} + \ket{\downarrow\uparrow} ) \Big].
\end{multline}
The spin correlations between $ss$ ($\bar{s}\bar{s}$) and $s\bar{s}$ are $P^{(2)}_{ss(\bar{s}\bar{s}),(b)}  = \frac{1}{3}$ and $P^{(2)}_{s\bar{s},(b)} = -\frac{2}{3}$, respectively.

Different color configurations give rise to distinct patterns of spin correlation. In general, the state of two-pair channel is expected to contain mixtures of both color configurations. For detailed calculations of the density matrices and the corresponding spin correlations, see Appendix.

\textit{\textbf{Decoherence: the Witness Effect.}}~In a Gedanken double-slit experiment, particles passing through the slits without observation would produce the well-known interference pattern on the screen; once an observer monitors which slit each particle traverses, the interference disappears and only two classical strips remain~\cite{Schlosshauer:2019ewh}. Monitoring of the system by the environment thus destroys quantum coherence.

Here we take the $s\bar{s}$ pairs as the system of interest and the other particles (partons or hadrons) produced during string breaking as the environment. Immediately after the vacuum excitation, the strange-quark pair is described by
\begin{equation}
    \ket{\text{sys}} = \ket{\psi_1} + \ket{\psi_2},
\end{equation}
where $\ket{\psi_{1(2)}}$ denote the spin components of the entangled state. At this stage, environmental degrees of freedom are absent, but they progressively emerge through successive string breakings. Since the color string is itself a highly entangled quantum state~\cite{Grieninger:2025rdi}, distinct environmental states are expected to couple to distinct system states, evolving the combined state into,
\begin{equation}
    \ket{\text{sys} \otimes\text{env} } = \ket{\psi_1}\ket{E_1} + \ket{\psi_2}\ket{E_2},
\end{equation}
where $\ket{E_{1(2)}}$ is the environmental state coupled to $\ket{\psi_{1(2)}}$. The environment can be described in either a partonic or a hadronic basis, complementary representations connected through nonperturbative QCD dynamics, and we adopt the hadronic basis, as hadrons are what experiments measure. The environmental state then decomposes into a direct product of individual hadron states,
\begin{equation}
    \ket{E_{1(2)}} = \left( \prod_{i=1}^{n}\otimes \right) \ket{e_{1(2)}^{(i)}},
\end{equation}
where $\ket{e_{1(2)}^{(i)}}$ is the state of the $i$-th hadron coupled to $\ket{\psi_{1(2)}}$. Tracing over the environment yields a reduced density matrix for the system whose off-diagonal elements are suppressed by a decoherence factor
\begin{equation}
    \alpha_n = \braket{E_1|E_2} = \prod_{i=1}^n  \braket{e_1^{(i)}|e_2^{(i)}},
\end{equation}
where $n$ is the number of environmental hadrons. Since in the SU(6) quark model~\cite{su6:model} the $\Lambda$ spin is fully carried by its strange quark, the spin correlation of the parent strange-quark pairs is directly accessible through the $\Lambda$ pairs. 

As a phenomenological approximation, each environmental hadron is assumed equally capable of distinguishing the system states, $\braket{e_1^{(i)}|e_2^{(i)}} = \beta$ for all $i$, giving $\alpha_n = \beta^n$. We further assume that the number of environmental hadrons within the $\Delta R$ interval between the $\Lambda$ pair follows a Poisson distribution, $p(n) = \frac{\lambda^n}{n!}e^{-\lambda}$ with mean $\lambda$. The averaged decoherence factor is then,
\begin{equation}
    \alpha (\Delta R) = \sum_{n} \beta^n\,p(n) = \exp\left[-(1-\beta)\,\lambda\right],
\end{equation}
with $\lambda = k\,\Delta R$, where $k$ is the average number of environmental hadrons per unit $\Delta R$. The observed spin correlation of the $\Lambda$ pair is therefore,
\begin{align}
     P(\Delta R ) = \alpha(\Delta R)\, P = P e^{-k^*\Delta R},
     \label{eq:P_DeltaR}
\end{align}
where $P$ is the initial spin correlation at production and $k^* = (1-\beta)k$. Since the multiplicity of environmental hadrons has not been directly measured by STAR or CMS, we use the charged-hadron yield per unit pseudorapidity as a proxy for scaling the environment size,
\begin{equation}
    k^{*}\propto \frac{dN}{d\eta},
    \label{eq:scale}
\end{equation}
which enables a prediction from RHIC to LHC energies.

Combining the VSC and the decoherence model, three free parameters are used to describe the data~\cite{STAR:2025njp,CMS_preliminary_result}: (i) $F$, the relative fraction of the one-pair channel with respect to the two-pair channel; (ii) $f$, the fraction of the color-symmetric configuration in the two-pair channel; and (iii) $k^{*}$, the strength of the decoherence effect, as summarized in Tab.~\ref{tab:parameters}. Both (i) and (iii) can in principle be independently measured, providing an independent validation of the model. The spin correlation between hyperon pair $\Lambda_1\Lambda_2$ is given by  
\begin{align}
  &  P_{\Lambda_1\Lambda_2} (\Delta R )=  A_{fd} \times \nonumber\\
    &~~~\left[ F\cdot P_{s_1 s_2}^{(1)} + (1-F) \cdot\big(f\cdot P^{(2)}_{s_1 s_2,(a) } + (1-f)\cdot P^{(2)}_{s_1 s_2,(b)}   \big)  \right]\nonumber\\
    &~~~\times \exp ( -k^* \Delta R   ) 
\end{align}
where $\Lambda_i\in\{\Lambda,\bar{\Lambda}\}$ $(i=1,2)$, and similar for strange quarks. Here $A_{fd}$ is the dilution factor accounting for feed-down effect, which is based on an estimate from the PYTHIA~8~\cite{Bierlich:2022pfr} simulation. We find that the feed-down fraction does not change significantly from RHIC to LHC energies, so the corresponding dilution factor remains approximately unchanged. We therefore take $A_{fd} = 1/3$ as suggested in Ref.~\cite{STAR:2025njp}. For $\Lambda \Lambda (\bar{\Lambda}\bar{\Lambda})$, the $F$ is taken to be 0, as one-pair channel does not contribute. 

\begin{table*}[htbp]
\centering
\caption{Model parameters with their definitions and values extracted from the STAR data~\cite{STAR:2025njp} and preliminary CMS data~\cite{CMS_preliminary_result}. }
\label{tab:parameters}
\begin{tabular}{c|c|c|c}
\hline\hline
Parameter & Definition & STAR & CMS \\
\hline
One-/two-pair fraction, $F$
& $N_{s\bar{s}}/(N_{s\bar{s}}+N_{ss\bar{s}\bar{s}})$
& $1.0\pm0.72$ & $0.0\pm0.12$ \\
Color-configuration fraction, $f$
& $N_{\ket{ss\bar{s}\bar{s};6_c\otimes\bar{6}_c}}/(N_{\ket{ss\bar{s}\bar{s};6_c\otimes\bar{6}_c}}+N_{\ket{ss\bar{s}\bar{s};\bar{3}_c\otimes3_c}})$
& $0.93\pm0.51$ & $0.78\pm0.05$ \\
Decoherence factor, $k^\ast$
& Strength of the decoherence effect
& $0.53\pm0.29$ & $1.29\pm0.737$ (fixed para.) \\
\hline\hline
\end{tabular}
\label{tab:parameters}
\end{table*}

 \begin{figure}[t]
    \centering
    \includegraphics[width=0.5\textwidth]{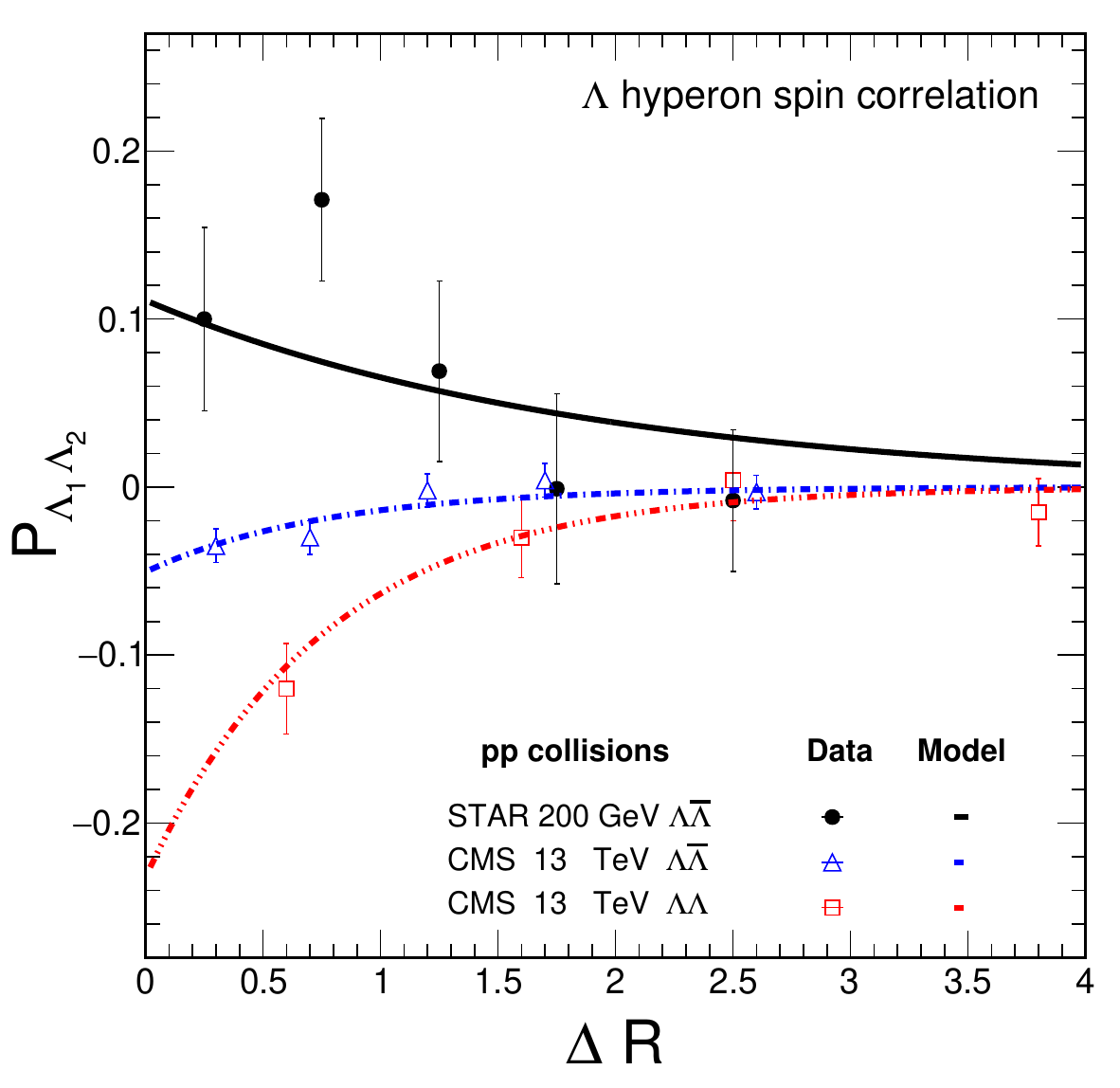}
    \caption{Spin correlation of $\Lambda$ hyperon pairs in $pp$ collisions of 200 GeV and 13 TeV measured at STAR and CMS experiment, respectively. The STAR data are taken from~\cite{STAR:2025njp}, and the CMS preliminary data are taken from~\cite{CMS_preliminary_result}. For STAR data, systematic uncertainties are added in quadrature with statistical uncertainties, while only statistical uncertainties are shown for CMS preliminary data.  }
    \label{fig:model}
\end{figure}

\textit{\textbf{Results.}}~Our strategy is first to perform a fit with all three parameters to the STAR $\Lambda\bar{\Lambda}$ spin correlation data~\cite{STAR:2025njp}. The extracted decoherence factor $k^*_{\text{RHIC}}$ is then used to predict the decoherence factor at LHC energy using the scaling relation in Eq.~\ref{eq:scale} as
\begin{equation}
    k_{\text{LHC}}^* = k_{\text{RHIC}}^* \frac{dN/d\eta |_{\eta=0,\text{LHC}}}{dN/d\eta|_{\eta=0,\text{RHIC}}},
    \label{eq:k_LHC}
\end{equation}
where $dN/d\eta|_{\eta=0,\text{LHC(RHIC)}}$ is the charged hadron yields at LHC (RHIC) energy~\cite{PHOBOS:2010eyu,CMS:2015zrm}. As a result, only two free parameters remain for the description of CMS preliminary data~\cite{CMS_preliminary_result}. Then, a simultaneous fit to the $\Lambda\bar{\Lambda}$ and $\Lambda\Lambda$ data is performed. The results are given in Fig.~\ref{fig:model}, with extracted parameters summarized in Tab~\ref{tab:parameters}. Notably, the decoherence factor $k^*$ at the LHC energy, extrapolated from RHIC through Eq.~\eqref{eq:k_LHC}, quantitatively reproduces the observed decoherence rate with increasing $\Delta R$, supporting the scaling behavior in Eq.~\eqref{eq:scale}. We find that this model can describe simultaneously the $\Lambda\bar{\Lambda}$ and $\Lambda\Lambda$ spin correlation data at two very different energies, providing the first experimental signatures on quarks spin entanglement in the vacuum and their quantum decoherence during hadronization. 

The extracted parameter $F$, although subject to large uncertainties, suggests that the one-pair channel contributes significantly at $\sqrt{s}=200~\mathrm{GeV}$, while the two-pair channel becomes increasingly important at $\sqrt{s}=13~\mathrm{TeV}$, indicating that higher-energy collisions favor the coherent excitation of multiple strange-quark pairs, consistent with the enhanced multi-strange hadron production observed at the LHC~\cite{ALICE:2016fzo}. With the one-pair channel dominating at RHIC, the color-configuration fraction $f$ is constrained only by the CMS data, which indicate a mixture of two color configurations in the two-pair channel, dominated by the color-symmetric state in Eq.~\eqref{eq:color_symmetric}. This can be understood from the one-gluon-exchange potential~\cite{mw52-xl56}: the quark-quark interaction is repulsive in the color-symmetric sextet ($6_c$) but attractive in the antitriplet ($\bar{3}_c$), so sextet quarks tend to hadronize into separate hadrons, whereas antitriplet quarks form a compact diquark and enter the same hadron. Alternative fits with different numbers of free parameters are presented in the Appendix. 

\textit{\textbf{Discussion.}}~While other mechanisms, such as gluon splitting and dihadron fragmentation~\cite{vonKuk:2025kbv,Zhang:2023ugf}, may in principle also generate spin correlations, the observed energy dependence and the interplay between spin-singlet and spin-triplet configurations in $\Lambda\Lambda$ and $\Lambda\bar{\Lambda}$ are difficult to reproduce through these effects; the STAR and CMS data together thus strongly support our proposed framework. 

Moreover, the framework yields distinctive predictions. First, the spin state in the one-pair channel is fully constrained by the vacuum quantum numbers, Eq.~\eqref{eq:ssbar}, which is a Bell state, exhibiting genuine quantum entanglement; therefore, future measurements of the $\Lambda\bar{\Lambda}$ spin density matrix would directly probe the degree of quantum entanglement. Second, the scaling relation in Eq.~\eqref{eq:scale} determines the decoherence factor at any collision energy once fixed at one; having passed the nontrivial test from RHIC to LHC energies, it can be further tested at other energies and at the future Electron-Ion Collider. Third, since the decoherence arises from environmental degrees of freedom generated during hadronization, the spin correlation may remain significant even at large $\Delta R$ if the hadron multiplicity is low, another direct test of the proposed mechanism based on \textit{the Witness Effect}.

Beyond this minimal one-pair plus two-pair framework, higher-pair excitations, nonzero orbital angular momenta, and mixed-flavor configurations (e.g., $s\bar{s}u\bar{u}$) may also contribute and are left for future investigation.

\textit{\textbf{Summary.}}~We develop a novel phenomenological framework to understand the spin correlation of $\Lambda$ hyperon pairs at RHIC and LHC energies. Specifically, the correlation originates from quark-antiquark pairs excited from the QCD vacuum, while the subsequent hadronization process induces quantum decoherence of the initial spin state. The model provides a remarkable description of STAR and CMS measurements. This work opens a new avenue for exploring the QCD vacuum and hadronization from the perspective of quantum information science.

\begin{acknowledgments}
The authors thank Dmitri Kharzeev for valuable discussions on QCD vacuum physics, Arjun Kumar and Li Xu for comments on the manuscript, and the CMS collaboration for discussing the preliminary data. We also acknowledge useful informal discussions with members of the Center for Frontier Nuclear Science (CFNS) at Stony Brook University and the local ePIC group at Brookhaven National Laboratory. In particular, F. Liu thanks CFNS for support through the 2026 Summer School, where part of this work was presented and completed. The work of F. Liu and Z. Tu was supported by the U.S. Department of Energy under Award No. DE-SC0012704 and by the Brookhaven National Laboratory Laboratory Directed Research and Development (LDRD) Project 26-029.
\end{acknowledgments}

\appendix
%
%
%
%
%
%
%


\maketitle
\section{Density Matrix of a Two-Qubit System}
The general density matrix of a system composed of two qubits, labeled \(a\) and \(b\), can be written as
\begin{align}
    \hat{\rho}= \frac{1}{4}
    \left(
        I_2\otimes I_2
        + B_i^{(a)}\,\sigma_i\otimes I_2
        + B_i^{(b)}\,I_2\otimes\sigma_i
        + C_{ij}\,\sigma_i\otimes\sigma_j
    \right),
    \label{eq:2qubit}
\end{align}
where \(I_2\) denotes the \(2\times2\) identity matrix and
\(\sigma_i\) denotes the \(i\)-th Pauli matrix. The indices
\(i,j\in\{x,y,z\}\) label the spatial directions, and repeated indices
are implicitly summed over. The coefficients \(B_i^{(a)}\) and
\(B_i^{(b)}\) characterize the spin polarizations of qubits \(a\) and
\(b\), respectively, along the \(i\)-th spatial direction, while
\(C_{ij}\) characterizes the spin correlation between the \(i\)-th
spin component of qubit \(a\) and the \(j\)-th spin component of
qubit \(b\). In particular,
\begin{equation}
    C_{ij} = \operatorname{Tr} \left[ (\sigma_i \otimes \sigma_j) \hat{\rho}  \right].
\end{equation}
A pair of quarks or \(\Lambda\) hyperons can naturally be regarded as
a two-qubit system, with their spin degrees of freedom fully described by the density matrix in Eq.~\eqref{eq:2qubit}. In addition, we introduce a rotationally-invariant spin-correlation variable measured in Ref.~\cite{STAR:2025njp,CMS_preliminary_result},
\begin{equation}
    P_{ab}
    =\operatorname{Tr}\left( 
    \frac{\vec{S}_a\cdot\vec{S}_b}
    {|\vec{S}_a|\,|\vec{S}_b|} \hat{\rho}\right)
    \label{eq:Pab_operator} 
\end{equation}
where $\vec{S}_{a(b)} = \frac{\hbar}{2}\vec{\sigma}_{a(b)}$, $|\vec{S}_{a(b)}|^2=\frac{3}{4}\hbar^2.$ The two variables of spin correlation are related by 
\begin{equation}
    P_{ab} = \frac{1}{3}\left( C_{xx} + C_{yy} + C_{zz} \right).
    \label{eq:Pab}
\end{equation}

\section{Vacuum Excitation}
\subsection{One-Pair Channel}
As demonstrated in the main text, the wave function of the one-pair channel is uniquely determined by the quantum numbers of the vacuum and found to be a Bell state. Here, we provide further details of the argument showing that this state is maximally entangled. We begin with a Bell state defined with respect to the spin-quantization axis $\hat{n}$,
\begin{align}
    \ket{\psi^{+},\hat{n}} = \frac{1}{\sqrt{2}} \left( \ket{\uparrow\downarrow,\hat{n}} + \ket{\downarrow\uparrow,\hat{n}} \right),
    \label{eq:Bell_n}
\end{align}
where $\hat{n} = (\sin\theta\cos\phi, \sin\theta\sin \phi,\cos\theta )$.  The spin eigenstates along $\hat{n}$ can be expanded in the basis quantized along the $z$ axis as  
\begin{align}
\ket{\uparrow,\hat{n}}& =  e^{-i\frac{\phi}{2} } \cos \frac{\theta}{2}  \ket{\uparrow} + e^{i \frac{\phi}{2} } \sin \frac{\theta}{2} \ket{\downarrow} \nonumber,\\
\ket{\downarrow,\hat{n}}& = -e^{-i \frac{\phi}{2}}\sin\frac{1}{2} \theta  \ket{\uparrow} + e^ {i \frac{\phi}{2}}\cos \frac{\theta}{2}  \ket{ \downarrow }.
\end{align}
Substituting these expressions into the Bell state in Eq.~\eqref{eq:Bell_n} yields
\begin{align}
  & \ket{\psi^+,\hat{n}}  = -\frac{1}{\sqrt{2}}\big( e^{-i\phi} \sin \theta \ket{\uparrow\uparrow}  \nonumber \\
  &~~~~~~~~-\cos \theta ( \ket{\uparrow\downarrow} + \ket{\downarrow\uparrow} )  - e^{i\phi} \sin \theta \ket{\downarrow\downarrow}    \big). 
\end{align}
which coincides, up to a normalization constant, with the wave function $\ket{s\bar{s}}^{(1)}$ shown in the main text. That says, the wave function $\ket{s\bar{s}}^{(1)}$ is a Bell state with respect to the direction of $s$ quark momentum. The associated density matrix, in helicity basis of $s\bar{s}$, is 
\begin{equation}
    \rho_{s\bar{s}} ^{(1)} = \ket{s\bar{s}}^{(1)}\bra{s\bar{s}}^{(1)} = \frac{1}{2}\begin{pmatrix} 
0 & 0 & 0 & 0 \\
0 & 1 & 1 & 0 \\
0 & 1 & 1 & 0 \\
0 & 0 & 0 & 0 \\
\end{pmatrix},
\end{equation}
with spin correlation as defined in Eq.~\eqref{eq:Pab_operator} 
\begin{equation}
    P_{s\bar{s}}^{(1)} = \frac{1}{3}. 
\end{equation}
It is worth noting that fermion--antifermion pairs produced in the decay of a state with $J^{PC}=0^{++}$ exhibit the same spin correlations, such as in Higgs-boson~\cite{Altakach:2022ywa} and $\chi_{c}^0$-resonance decays~\cite{Fabbrichesi:2024rec}.

\subsection{Two-Pair Channel}
Here we provide more details on the spin correlation of two-pair channel. The $ss\bar{s}\bar{s}$ system is organized as diquark-antidiquark configuration, and wave function is decomposed into flavor, spatial, spin, and color components, i.e.,
\begin{equation}
    \ket{ss\bar{s}\bar{s}} = \ket{\text{flavor}}\otimes\ket{\text{spatial}}\otimes \ket{\text{spin}}\otimes\ket{\text{color}}.
\end{equation}
We consider each component separately. 
\begin{itemize}
    \item The flavor wave function is symmetric for strange diquark (anti diquark). 
    \item Let $l_{ss}$ ($l_{\bar{s}\bar{s}}$) denote the internal orbital angular momentum within the diquark (antidiquark), and $L$ the orbital angular momentum between the diquark and antidiquark. The lowest-energy configuration is obtained by setting $l_{ss}=l_{\bar{s}\bar{s}} = L =0$, making the state easiest to excite from vacuum. The parity is thus $P= \eta_s^2\eta_{\bar{s}}^2 = +1 $. 
    \item The color configuration of a diquark can be either antisymmetric color-antitriplet state ($\bar{3}_c$), or symmetric color-sextet state ($6_c$). For anti-diquark, the color configuration can be either color-triplet state ($3_c$), or color-antisextet ($\bar{6}_c$). 
    \item The (anti) diquark spin wave function can be either in the symmetric spin-triplet configuration or the antisymmetric spin-singlet configuration. In both cases, the diquark and antidiquark spins are required to couple to total spin $S=0$, which leads to $J=0$ in the absence of orbital angular momentum.
\end{itemize}

Therefore, two possible spin configurations are allowed. For each configuration, the spin correlation can be extracted with the following procedures.  First, one can begin with the density matrix constructed from the wave function $\ket{ss\bar{s}\bar{s}}^{(2)}$, i.e.,  
\begin{equation}
    \rho_{ss\bar{s}\bar{s}}^{(2)} = \ket{ss\bar{s}\bar{s}}^{(2)} \bra{ss\bar{s}\bar{s}}^{(2)}.
\end{equation}
Then, reduced density matrix of $ss$ is obtained by tracing out the degrees of freedom of $\bar{s}\bar{s}$ , i.e., 
\begin{equation}
    \rho_{ss}^{(2)} = \text{Tr}_{\bar{s}\bar{s}} \rho_{ss\bar{s}\bar{s}}^{(2)}. 
\end{equation}
The reduced density matrix of $\bar{s}\bar{s}$ and $s\bar{s}$ can be found similarly. The spin correlation is eventually extracted through Eq.~\eqref{eq:Pab_operator}. The reduced density matrix and spin correlation are summarized below. 

\textbf{Configuration (a)}: If the $ss$ ($\bar{s}\bar{s}$) is the symmetric color-(anti) sextet state, the spin state has to be in the spin singlet. The spin wave function is reproduced from the main text for completeness  
\begin{equation}
    \ket{ss\bar{s}\bar{s};6_c\otimes\bar{6}_c}^{(2)} = \frac{1}{2} \left( \ket{\uparrow\downarrow}-\ket{\downarrow\uparrow}\right)\left(\ket{\uparrow\downarrow} - \ket{\downarrow\uparrow} \right).
    \label{eq:color_symmetric_s}
\end{equation}
The reduced density matrix of  $ss$ ($\bar{s}\bar{s}$) is,  
\begin{equation}
    \rho^{(2)}_{ss(\bar{s}\bar{s}),(a)} = \frac{1}{2}\begin{pmatrix} 
0 & 0 & 0 & 0 \\
0 & 1 & -1 & 0 \\
0 & -1 & 1 & 0 \\
0 & 0 & 0 & 0 \\
\end{pmatrix},
\end{equation}
The spin correlation between  $ss$ ($\bar{s}\bar{s}$) is, 
\begin{equation}
    P^{(2)}_{ss(\bar{s}\bar{s}),(a)} = -1. 
\end{equation}

The reduced density matrix of  $s\bar{s}$ is:  
\begin{equation}
    \rho^{(2)}_{s\bar{s},(a)} = \frac{1}{4}\begin{pmatrix} 
1 & 0 & 0 & 0 \\
0 & 1 & 0 & 0 \\
0 & 0 & 1 & 0 \\
0 & 0 & 0 & 1 \\
\end{pmatrix},
\end{equation}
Then spin correlation between $s\bar{s}$ is, 
\begin{equation}
    P^{(2)}_{s\bar{s},(a)} = 0. 
\end{equation}

\textbf{Configuration (b)}: If the $\bar{s}\bar{s}$  $(ss)$ is in the antisymmetric color-(anti) triplet state, the spin state is forced to be in the spin triplet. The spin wave function is reproduced from the main text 
\begin{multline}
\ket{ss\bar{s}\bar{s};\bar{3}_c\otimes3_c}^{(2)} = \frac{1}{\sqrt{3}} \ket{S=1}_{ss}\ket{S=-1}_{\bar{s}\bar{s}} \\
- \frac{1}{\sqrt{3}} \ket{S=0}_{ss}\ket{S=0}_{\bar{s}\bar{s}} + \frac{1}{\sqrt{3}} \ket{S=-1}_{ss}\ket{S=1}_{\bar{s}\bar{s}} \\
= \frac{1}{\sqrt{3}} \Big[ \ket{\uparrow\uparrow}\ket{\downarrow\downarrow} + \ket{\downarrow\downarrow}\ket{\uparrow\uparrow} \\
- \tfrac{1}{2} ( \ket{\uparrow\downarrow} + \ket{\downarrow\uparrow} ) ( \ket{\uparrow\downarrow} + \ket{\downarrow\uparrow} ) \Big].
\end{multline}
The reduced density matrix of  $ss$ ($\bar{s}\bar{s}$) is
\begin{equation}
    \rho_{ss(\bar{s}\bar{s}),(b)} = \begin{pmatrix} 
1/3 & 0 & 0 & 0 \\
0 & 1/6 & 1/6 & 0 \\
0 & 1/6 & 1/6 & 0 \\
0 & 0 & 0 & 1/3\\
\end{pmatrix}.
\end{equation}
The spin correlation between $ss$ ($\bar{s}\bar{s}$) is 
\begin{equation}
    P_{ss(\bar{s}\bar{s}),(b)}  = \frac{1}{3}.
\end{equation}
The reduced density matrix of  $s\bar{s}$ is
\begin{equation}
    \rho_{s \bar{s},(b)} = \begin{pmatrix} 
1/12 & 0 & 0 & 0 \\
0 & 5/12 & -1/3 & 0 \\
0 & -1/3 & 5/12 & 0 \\
0 & 0 & 0 & 1/12 \\
\end{pmatrix}.
\end{equation}
The spin correlation between $s\bar{s}$ is
\begin{equation}
    P_{s\bar{s},(b)} = -\frac{2}{3}.
\end{equation}

\section{Reducing Free Parameters}
\begin{figure*}[t]
    \centering
    \includegraphics[width=1\textwidth]{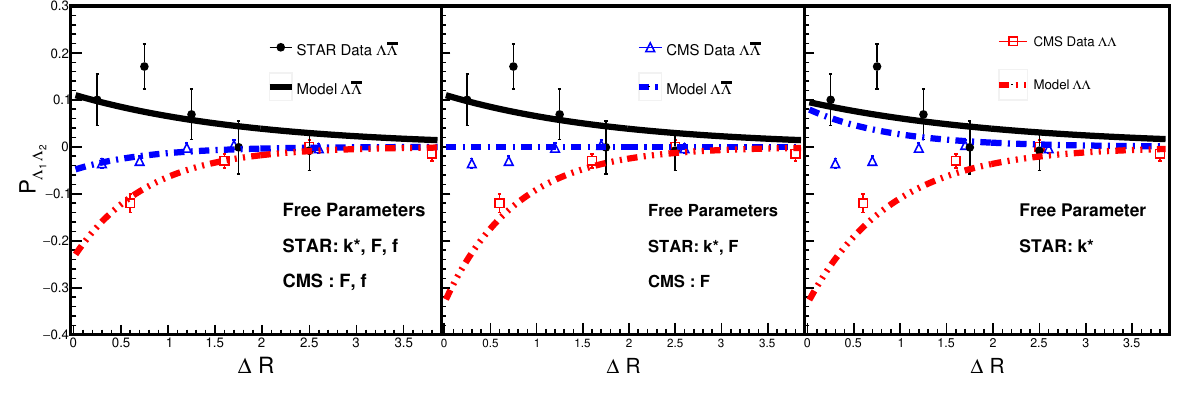}
    \caption{A sequence of fits with progressively decreasing free parameters from left to right panels.  The filled markers represent the STAR data taken from~\cite{STAR:2025njp}, and the open markers represent the CMS preliminary data taken from~\cite{CMS_preliminary_result}. For STAR data, systematic uncertainties are added in quadrature with statistical uncertainties, while only statistical uncertainties are shown for CMS preliminary data.}
    \label{fig:fit_seq}
\end{figure*}

To further assess the capability of the proposed framework in describing the experimental data, a sequence of fits with progressively decreasing free parameters is performed as shown in Fig.~\ref{fig:fit_seq}. The full three-parameter fit described above is shown in the left panel for comparison. As discussed in the main text, the symmetric color configuration is more likely to contribute to the di-hyperon spin correlation.  Motivated by this, we set $f=1$, ignoring the contribution from anti-symmetric color configuration completely. The STAR $\Lambda\bar{\Lambda}$ data are then fitted with two free parameters. Then extracted decoherence $k^*$ is extrapolated to the LHC energy using relation $k^*\propto \frac{dN}{d\eta}$, leaving one free parameter $F$ for simultaneous fit to both CMS $\Lambda\Lambda$ and $\Lambda\bar{\Lambda}$ data. The result is shown in the middle panel of Fig.~\ref{fig:fit_seq}. 
Furthermore, the relative fraction of one-pair and two-pair channels can be estimated phenomenologically from the measured strange-hadron yields. The rapidity-differential yield of $s\bar{s}$ pair can be calculated as: 
\begin{align}
    \frac{d N(s\bar{s}) }{d y}|_{y= 0}  = \frac{1}{2} (  2 \frac{dN(K_s^0)}{dy} + \frac{dN(K^++K^-)}{dy}  \nonumber
    \\+ \frac{dN(\Lambda +\bar{\Lambda})}{dy} + 2\frac{dN(\Xi + \bar{\Xi})}{dy} +3\frac{dN(\Omega^- + \bar{\Omega}^+)}{dy}    )|_{y=0} 
\end{align}
For the pair multiplicity distribution, we employ a Poisson distribution, 

\begin{equation}
    p(n_{s\bar{s}}) = \frac{\mu^{n_{ss}} }{n_{s\bar{s}}!} e^{-\mu}, 
\end{equation}
where $n_{s\bar{s}}$ is the number of $s\bar{s}$ produced per unit of rapidity, and $\mu$ is the average of $n_{s\bar{s}}$ with $\mu =  dN(s\bar{s})/dy|_{y=0}$.  The fraction of one-pair channel is thus found to be: 
\begin{equation}
    F = \frac{p(n_{s\bar{s}}  =1 )}{p( n_{s\bar{s}}  =1 ) +p( n_{s\bar{s}}  =2) }. 
\end{equation}
The Poisson distribution naturally describes independent pair production. Although the two-pair excitation is not genuinely independent process, the Poisson distribution can be used as a baseline of independent pair creation. In this sense, the suppression or enhancement may point to the intrinsic property of vacuum excitation dynamics. 

With parameter $F$ constrained by this estimate, the fit to the STAR data contains only one remaining free parameter, while the CMS data of $\Lambda\Lambda$ and $\Lambda\bar{\Lambda}$ spin correlation are described without any additional free parameters. The result is shown in the right panel of Fig.~\ref{fig:fit_seq}.

\begin{table}
    \caption{Strange hadron yields $\frac{dN}{dy}$ ~\cite{STAR:2006nmo,ALICE:2020jsh}  in $pp$ collisions at $\sqrt{s}$= 200~GeV and 13~TeV. }
    \centering
    \begin{tabular}{ccc}
        \hline\hline
                 &      $pp$ 200 GeV& $pp$ 13 TeV\\
        \hline
       $K_S^0$  & 0.134 $\pm$ 0.011& 0.3192 $\pm$ 0.0111\\
       $K^++K^-$ & 0.277 $\pm$ 0.014& 0.6205 $\pm$ 0.0303\\
       $\Lambda +\bar{\Lambda}$ & 0.0834 $\pm$ 0.0056& 0.1807 $\pm$ 0.0102\\
       $\Xi +\bar{\Xi}$ & 0.0055 $\pm$ 0.0014&   0.01980 $\pm$ 0.00083\\
       $\Omega^- + \bar{\Omega}^+$  &    0.00034 $\pm$ 0.00019& 0.001846 $\pm$ 0.00013\\     
        $s + \bar{s}$ &   0.64 $\pm$ 0.019&      1.48 $\pm$ 0.15\\
        \hline\hline
        \end{tabular}
    
    \label{tab:dndy}
\end{table}

\clearpage

\bibliography{reference}

\end{document}